\documentclass{llncs}
\usepackage{amsmath,amssymb,amsfonts}
\usepackage{mathtools}
\usepackage[usenames,dvipsnames]{color}
\usepackage{xcolor}
\usepackage{xspace}
\usepackage{microtype}
\usepackage{todonotes}
\usepackage{colonequals}
\usepackage{marvosym}
\usepackage{wasysym}
\usepackage{booktabs}
\usepackage{refcount}
\usepackage{makecell}

\usepackage{multirow}
\usepackage{multicol}

\usepackage[T1]{fontenc}
\usepackage{paralist}

\usepackage{subfigure}
\usepackage{url}
\usepackage{appendix}
\usepackage{graphicx}
\usepackage{float}
\usepackage{algorithm}
\usepackage{listings}
\usepackage{newalg}
\usepackage{nicefrac}
\usepackage{tikz} 
\usepackage{pgfplots}
\usepackage{marvosym}
\usepackage{wrapfig}

\usepackage{footnote}
\usepackage{tablefootnote}
\usepackage{array}

\usetikzlibrary{arrows,decorations.pathmorphing,positioning,fit,trees,shapes,shadows,automata,calc} 

\usepackage{pdfpages}

\usepackage{preamble/macros} 

\begin{document}
  \title{The Probabilistic Model Checker \storm}
\subtitle{(Extended Abstract)}
\author{Christian Dehnert \and Sebastian Junges \and Joost-Pieter Katoen \and Matthias Volk}

\institute{RWTH Aachen University}
\vspace{1.4cm}
\maketitle

  \noindent We present a new probabilistic model checker \storm.
Using state-of-the-art libraries, we aim for both high performance and versatility.
This extended abstract gives a brief overview of the features of \storm.
  \paragraph{Goals.}
Probabilistic model checking has matured immensely in the past decade \cite{DBLP:conf/sfm/KwiatkowskaNP07,DBLP:conf/lics/Kwiatkowska03,DBLP:series/natosec/Baier16,Kat16}.
Applications and uses go far beyond the standard algorithms and objectives.
We therefore developed an extensible toolbox for probabilistic model checking that offers reusable modules to quickly implement new functionality.
The rise of more and more specialized tools witnesses that a single tool is rarely suitable for the broad range of applications and models.
Consequently, we offer a range of engines that make use of various backend solvers.
This core is aiming for high performance and low memory requirements.
On top of it we realized sophisticated techniques ranging from generation of probabilistic counterexamples \cite{DBLP:conf/atva/DehnertJWAK14}, over permissive scheduler generation \cite{junges-et-al-tacas-2016} and parameter synthesis \cite{dehnert-et-al-cav-2015,DBLP:conf/atva/QuatmannD0JK16} to the analysis of dynamic fault trees \cite{DBLP:conf/safecomp/0001JK16} and probabilistic program verification \cite{DBLP:conf/atva/0001DKKW16}.
Finally, the wide range of input languages supports the development of new techniques in a variety of fields.
\paragraph{Supported model types.} \storm{} supports both discrete-time and con\-tin\-u\-ous-time Markov models and nondeterministic variants thereof. 
Table~\ref{tab:modeltypes} gives an overview and classification about the concrete model types: discrete-time Markov chains (DTMCs), continuous-time Markov chains (CTMCs), Markov decision processes (MDPs) and Markov automata (MA), the compositional variant of continuous-time Markov Decision processes (CTMDPs) \cite{Kat16}.
Additionally, all these model types can be enriched with reward models.
\begin{table}
  \def\arraystretch{1.3}\tabcolsep=10pt
  \begin{center}
    \caption{Model types supported by \storm.}
    \begin{tabular}{ccc}
      \toprule
      & discrete-time & continuous-time \\
      \cmidrule{1-3}
      deterministic & DTMCs & CTMCs \\
      nondeterministic & MDPs & MA \\
      \bottomrule
    \end{tabular} 
    \label{tab:modeltypes}
  \end{center}
\end{table}
\paragraph{Modelling languages.}
There are several modelling languages that can be used to specify the aforementioned model types. 
First of all, in the spirit of MRMC \cite{mrmc}, models can be passed in a format that explicitly enumerates transitions. 
However, \storm{} also supports a variety of symbolic input formats. 
Most prominently, the \prism{} input language \cite{prism_website} is supported, which enables us to consider all benchmark models from the \prism{} benchmark suite \cite{KNP12b} that represent supported model types. 
In an attempt to unify the probabilistic modelling language landscape, the JANI format \cite{jani} is currently under development and its first version is included in \storm.
Besides, it is the first tool to support \emph{every} generalized stochastic Petri net (GSPN) \cite{DBLP:conf/apn/EisentrautHK013} via both a dedicated model builder as well as an encoding in JANI.
Additionally, \storm{} features the analysis of dynamic fault trees (DFTs) \cite{handbook2002} and has been shown to outperform competing tools in this domain \cite{DBLP:conf/safecomp/0001JK16}.
Finally, we support probabilistic programs written in the conditional probabilistic guarded command language (cpGCL) \cite{DBLP:series/mcs/McIverM05}, again via a reduction to JANI.

\paragraph{Properties.}
The main focus of our tool is probabilistic branching time logics, i.e.~PCTL \cite{HJ94} and CSL \cite{DBLP:conf/cav/AzizSSB96} for discrete-time and continuous-time models, respectively. 
To enable the treatment of reward-objectives, we support reward extensions of these logics in a similar way as PRISM.
As conditional probabilities and conditional rewards \cite{baier_cond} have proven to express interesting properties, we also support these.
%
\paragraph{Engines.}
There are different representations of probabilistic models in memory that differ in efficiency depending on the characteristics of the input model.
\storm{} features two distinct representations: while \emph{sparse matrices} tend to work well for small and moderately sized models, multi-terminal decision diagrams (MTBDDs) are able to represent gigantic systems.
To enable the treatment of a broader class of input models, the user can select between several engines built around the two in-memory representations to perform the verification tasks.
Both \storm's \textit{sparse} and the \textit{learning} engine purely use a sparse matrix based representation. While the former amounts to an efficient implementation of the standard approaches, the latter one implements the ideas of \cite{DBLP:conf/atva/BrazdilCCFKKPU14} to provide sound statistical model checking for discrete-time models.
Two other engines, \textit{dd} and \textit{hybrid}, use MTBDDs as the primary representation.
While \textit{dd} exclusively uses decision diagrams, \textit{hybrid} also uses sparse matrices for operations deemed more suitable on this data format.
\paragraph{Solvers.}
\storm's infrastructure is built around the notion of a \emph{solver}. 
For instance, solvers are available for sets of linear or Bellman equations (both using sparse matrices as well as MTBDDs), (mixed-integer) linear programming (MILP) and satisfiability modulo theories (SMT) solving. 
Offering these interfaces has several key advantages.
First, it provides easy and coherent access to the tasks commonly involved in probabilistic model checking.
Secondly, it enables the use of dedicated state-of-the-art high-performance libraries for the task at hand. More specifically, as the performance characteristics of different backend solvers can vary drastically for the same input, this permits choosing the best solver for a given task.
Licensing problems are avoided, because implementations can be easily enabled and disabled, depending on whether or not the particular license fits the requirements.
Finally, implementing new solver functionality is easy and can be done without knowledge about the global code base.
For each of those interfaces, several actual implementations exist. Table~\ref{tab:solvers} gives an overview over the currently available implementations.
\begin{table}[t]
  \def\arraystretch{1.3}\tabcolsep=10pt
  \begin{center}
  
    \caption{Solvers offered by \storm.}
    \begin{tabular}{ll}
      \toprule
      solver type & available solvers \\
      \cmidrule{1-2}
      linear equations (sparse) & Eigen \cite{eigenweb}, gmm++ \cite{gmmpp_website}, built-in \\
      linear equations (MTBDD) & CUDD \cite{cudd_website}, Sylvan \cite{DBLP:conf/tacas/DijkP15} \\
      Bellman equations (sparse) & Eigen, gmm++, built-in \\
      Bellman equations (MTBDD) & CUDD, Sylvan \\
      (MI)LP & Gurobi \cite{gurobi}, glpk \cite{glpkweb} \\
      SMT & Z3 \cite{dMB08}, MathSAT \cite{DBLP:conf/tacas/CimattiGSS13}, SMTLIB \cite{Barrett10c.:the} \\
      \bottomrule
    \end{tabular} 
    \label{tab:solvers}
  \end{center}
\end{table}
Many components in \storm{} make use of these solvers. 
The most prominent example is the obvious use of the equation solvers for answering standard verification queries. 
However, various other modules use them too, for example model generation (SMT), counterexample generation \cite{DBLP:journals/corr/abs-1305-5055,DBLP:conf/atva/DehnertJWAK14} (SMT, MILP) and permissive scheduler generation \cite{DBLP:journals/corr/DragerFK0U15,junges-et-al-tacas-2016} (SMT, MILP).

\paragraph{Parametric models and exact arithmetic.}
\storm was used as backend in \cite{dehnert-et-al-cav-2015,DBLP:conf/atva/QuatmannD0JK16}. By using the dedicated library \tool{CArL} \cite{carl_website} for the representation of rational functions and applying novel algorithms for the analysis of parametric discrete-time models, it has proven to significantly outperform other tools. 
In addition, for these models \storm{} is able to compute exact solutions rather than using floating point arithmetic.

\paragraph{Usage.}
\storm{} can be used via three interfaces. 
For end-users we provide command-line interfaces: There are several binaries that provide specialized access to the available settings for different tasks. 
For example, \tool{storm-dft} offers the features and settings related to the analysis of DFTs. 
Advanced users can utilize one of the many settings to tune the performance.
Developers may either use a C++ API that offers fine-grained and performance-oriented access to \storm's functionality, or use a Python API which supports rapid prototyping and allows users to profit from high-performance implementations within \storm.

\paragraph{Related work.}
There are several tools whose functionality overlaps with that of \storm (in alphabetical order): \tool{FACT} \cite{DBLP:conf/tacas/CalinescuJP16}, \tool{FIG} \cite{FIGpaper}, \tool{IMCA} \cite{DBLP:conf/atva/GuckTHRS14}, \tool{iscasMC} \cite{iscasmc}, \tool{Modest} \cite{DBLP:conf/tacas/HartmannsH14}, \tool{MRMC} \cite{mrmc}, \tool{PARAM} \cite{PARAM10}, \tool{PASS} \cite{DBLP:conf/tacas/HahnHWZ10}, \tool{PAT} \cite{DBLP:conf/issre/LiuSD11}, \tool{PRISM} \cite{KNP11} and \tool{Uppaal SMC} \cite{DBLP:journals/corr/abs-1207-1272}. Moreover, there are specialized tools for the generation of probabilistic counterexamples (\tool{COMICS} \cite{jansen-et-al-atva-2012}), DFTs (\tool{DFTCalc} \cite{ABBGS13}) and GSPNs (\tool{GreatSPN} \cite{DBLP:conf/apn/AmparoreBD14}).
Despite the existence of these tools, we believe that \storm{} is unique in its trade-off between performance and modularity, the supported solvers and wide range of supported modelling languages.

  \pagebreak
  \bibliographystyle{splncs}
  \bibliography{literature}
\end{document}